\documentstyle[preprint,aps,epsf,graphicx]{revtex}
\begin{document}
\draft
%\twocolumn[
%\hsize\textwidth\columnwidth\hsize\csname @twocolumnfalse\endcsname

\draft
\title{
Crossover between Equilibrium and Shear-controlled Dynamics in
Sheared Liquids
}
\author{
        L.~Angelani$^{1,2}$,
        G.~Ruocco$^{1}$,
        F.~Sciortino$^{1,2}$,
        P.~Tartaglia$^{1,2}$,
        and F.~Zamponi$^{1}$
        }
\address{
         $^1${Dipartimento di Fisica and INFM, Universit\`a di Roma
         {\em La Sapienza}, P. A. Moro 2, 00185 Roma, Italy}
        }
\address{
        $^2${INFM - Center for Statistical Mechanics and Complexity,
        Universit\`a di Roma {\em La Sapienza}, P. A. Moro 2, 00185 Roma, Italy}
        }

\date{\today}
\maketitle
\begin{abstract}
We present a numerical simulation study of a simple monatomic
Lennard-Jones liquid under shear flow, as a function of both
temperature $T$ and shear rate $\dot\gamma$. 
By investigating different observables we find that 
{\it i)} It exists a line, $T_{\dot \gamma}$, 
in the ($T$-$\dot\gamma$) plane that sharply marks the boarder 
between an {\it ``equilibrium''} and a {\it ``shear-controlled''} region for 
both the dynamic and
the thermodynamic quantities; and 
{\it ii)} Along this line the
structural relaxation time, $\tau_\alpha(T_{\dot \gamma})$, is
proportional to $\dot\gamma^{-1}$, i.~e. to the typical
time-scale introduced by the shear flow. Above $T_{\dot \gamma}$
the liquid dynamics is unaffected by the shear flow, while below
$T_{\dot \gamma}$ both $T$ and $\dot \gamma$ control the particle
motion.
\end{abstract}
\pacs{PACS Numbers : 61.20.Lc, 64.70.Pf, 47.50.+d}
%83.10.Rs}
%02.70.Ns}
%]

%%%%%%%%%%%%%%%  TEXT  %%%%%%%%%%%%%%%%

The effects of the shear flow on the properties of simple
liquids has recently been the object of extensive investigations
\cite{utdest00,yaon,beba,stwe82,mela98,la01}. Beside its
intrinsic theoretical interest (the study of out-of-equilibrium
stationary states), it has been hypothesized
\cite{cukupe97,bebaku00} that the
shear flow acts as an aging-stopping mechanism, 
thus suggesting an interesting experimental method to study 
dynamics in supercooled liquids and glasses. Indeed, under
shear flow, the system is in an out-of-equilibrium situation but
reaches a stationary regime: similarly to equilibrium the
correlation functions depend only on the time-difference. 
This takes place also at those temperatures for which the non-sheared
system would be in an ${\it aging}$ regime (i.~e. the correlation
functions depend explicitly on two times). In the common phrasing,
the application of a shear flow has the effect to rejuvenate the
glassy system and the aging phenomenon characteristic of glasses
is stopped. The fact that some relevant properties of aging
systems, like the existence of a generalized
fluctuation-dissipation relation \cite{cukupe97,bocukume97} and
of an effective temperature, still hold in systems under shear
flow \cite{bebaku00,babe00}, makes the study of sheared systems 
an important topic.
This is particularly true in view of possible
experiments aiming to probe the out-of-equilibrium dynamics of
glassy systems: 
in aging the waiting-time dependence of correlation functions 
prevents the acquisition of the data with the desired statistics,
while under shear correlation functions become time translation invariant.

In this work we use numerical simulations to study the effects
of the shear flow on the properties of a simple
Lennard-Jones model liquid. In particular we compare the
temperature dependence of some physical quantities of the driven
system (with different shear rates $\dot{\gamma}$) with those of
the equilibrium system ($\dot{\gamma}=0$). We find that
{\it i)} it exists a well-defined crossover temperature
$T_{\dot\gamma}$, whose value depends on the shear rate
$\dot{\gamma}$, below which the properties of the sheared 
system exhibit a marked difference from
those of the equilibrium system. On the contrary, for
$T>T_{\dot\gamma}$ the driven system is not influenced by the
presence of a shear flow, and both the energy and the structural
relaxation time coincide with their equilibrium values.
{\it ii)} The same $T_{\dot\gamma}$ can be derived from the $T$-dependence
of both potential energy and structural relaxation times,
thus indicating the robustness of the "cross-over" temperature
concept; {\it iii)} At $T_{\dot \gamma}$, the structural
relaxation time $\tau_\alpha(T_{\dot \gamma})$ is proportional to
$\dot\gamma^{-1}$, i.~e. to the typical time-scale introduced by
the shear, the proportionality coefficient being an observable
dependent quantity. The previous observations lead to a microscopic
interpretation of the shear thinning effect, and suggest a
quantitative experimental test on the temperature dependence of
the non-linear viscosity in simple liquids.

The investigated system is made of $N$=256 particles interacting
via a simple Lennard-Jones potential, plus a small many-body term
\cite{dianparu00} introduced to prevent the crystallization 
unavoidably occurring in undercooled monatomic systems. 
The particles are confined in a cubic box, at
density $\rho = 1$ (hereafter all the quantities are expressed in
reduced LJ unit), with periodic boundary condition adapted to the
presence of a shear flow. The latter is applied to the system
along the $x$ direction with a gradient velocity field along the
$y$ axis. The molecular dynamics simulation is performed using
SLLOD algorithm \cite{evans}, with a Nose-Hoover thermostat for
the thermal velocities. Different shear rates $\dot{\gamma}$ were
studied in the range $\dot{\gamma}=10^{-1}\div\dot{\gamma}=
10^{-3}$. A preliminary simulation performed at $\dot{\gamma}=0$
was used to determine the reference behavior of the equilibrium
system. For all the shear rates considered, different physical
quantities are analyzed as a function of temperature: energy,
incoherent scattering functions, relaxation times.

In Fig. 1 we report the potential energy per particle $e$ as a function
of temperature for selected values of the shear rate. The full
line indicates the temperature dependence of $e$  at equilibrium 
($\dot{\gamma}=0$), which, as shown in
previous work \cite{dianparu00,copave00,sckota99}, can be described quite
accurately by the Rosenfeld-Tarazona $T^{3/5}$ power-law
\cite{rota98}. The open symbols indicates the caloric data for
three selected shear rates: $\dot{\gamma}$=1$\cdot$10$^{-1}$,
4$\cdot$10$^{-2}$, 6$\cdot$10$^{-3}$. At high temperature, one
can observe a good agreement between $e(T,\dot\gamma)$ and
$e(T,\dot\gamma=0) = e_{eq.}(T)$. On lowering the temperature, the agreement
breaks down at a $\dot\gamma$-dependent temperature
($T_{\dot\gamma}$), below which the caloric curve of the sheared
system deviate from the $T^{3/5}$ power-law. In particular, at
low-$T$, $e(T,\dot\gamma)$ is well described by a linear
temperature dependence (indicated, as an example, by the dashed
line for $\dot{\gamma}$=4$\cdot$10$^{-2}$). The deviation of the
energy of the driven system from the equilibrium energy is
evidenced in the inset of Fig. 1, where the energy difference
$e-e_{eq.}$ is reported as a function of temperature for the
system with $\dot{\gamma}=4\cdot10^{-2}$. The existence of a
linear behavior of $e(T,\dot\gamma)$ at low $T$ and of an
analytic expression for this quantity at high $T$ allows for a
straightforward identification of a crossover temperature
$T_{\dot\gamma}$ (indicated in the inset of Fig.~1
with dashed arrow).

In Fig. 2 we report the $\dot{\gamma}$-dependence of
the parameters describing the caloric curves of the sheared
systems: the crossover temperature $T_{\dot\gamma}$ and the
potential energy value at $T$=0 $e_0$. As shown in Fig.~2a,
$T_{\dot\gamma}$ is an increasing function of the shear rate. For
small shear rate values it approaches a {\it plateau} whose value
is very close to the estimated mode-coupling critical temperature
$T_{MCT}$ ($T_{MCT}$=0.475 for this potential model
\cite{mct,andiruscsc00}) indicated by the dashed line in Fig.~2a.
A similar {\it plateau} is observed in the $\dot{\gamma}$
dependence of $e_0$ (see Fig.~2b). Here, for small shear rate
values, $e_0$ reaches the value $e_0^{pl.} \sim -6.90$. We note
that the value $e_0^{pl.}$ is higher than the lowest inherent
structure energy value obtained in equilibrium simulation
$e_0^{IS}=-7.0$ \cite{desc}. Similarly to the case of analytic
mean field spin glass models \cite{bebaku00}, the {\it plateau}
value can be interpreted as a threshold in the potential energy
surface, above which the system is forced by the shear. However,
as we are dealing with a non-mean field system, we expect that
for values of $\dot\gamma$ small enough -well below the ones that
we are able to study- the system cross the threshold. 
Similarly, also in Fig.~2a the
existence of a {\it plateau} is only apparent and mirrors the
(apparent) power-law divergence of the relaxation times predicted
by the MCT. The evaluation of the crossover temperature for even
smaller ${\dot\gamma}$, i.~e. in a region not accessible to the
simulation for CPU time reason, would have resulted in a
$T_{\dot\gamma}$ smaller than $T_{MCT}$ as a consequence of the
presence of activated processes in the investigated non-mean field
system.

Further information on the effect of the shear flow can be
obtained analyzing the temperature dependence of correlation
functions in driven systems. For the different shear rates
considered, we have calculated the incoherent scattering
functions $F_{\bf q}(t)$:
\begin{equation}
F_{\bf q}(t) = \frac{1}{N} \sum_{j=1}^{N} \langle e^{i{\bf q} \cdot
[{\bf r}_j(t)-{\bf r}_j(0) ] } \rangle \ ,
\label{fq}
\end{equation}
where ${\bf r}_j(t)$ is the position of particle $j$ at time $t$.
The wave vector ${\bf q}$ is that of the first peak of
the static structure factor $S_{\bf q}$ ($q_{max}=7.1$) along the
``shear-free'' direction (${\bf q}=(0,0,q_{max})$). In Fig.~3 we
report the $F_{\bf q}$ for the equilibrium system (full lines)
and, as an example, for the system with
$\dot{\gamma}$=4$\cdot$10$^{-2}$ (dashed lines),
for three different temperatures.
At the higher reported temperature ($T=1.6$), the
$F_{\bf q}$ of the sheared system is undistinguished from the
equilibrium one. At the intermediate temperature ($T=0.80$) the
full and dashed lines start to deviate one from the other, an
effect that become more and more clear on lowering the
temperature ($T=0.56$). It is worth to note that -at the reported value of
$\dot\gamma$- the crossover temperature derived from
Fig.~1 is $T_{\dot\gamma}$=0.97, i.~e. intermediate between the
the first two reported $F_{\bf q}$. From the inspection of
Fig.~3, one can conclude that for $T$ smaller than 
$T_{\dot\gamma}$ (derived from $e$)
there are no effects of the shear flow on the dynamics.

To put the previous observation on a quantitative ground, we
analyze the temperature dependence of relaxation time 
$\tau_\alpha$, 
defined as the time at which $F_{\bf q}$ reach $1/e$-th of
its non-ergodicity factor (the apparent plateau value). 
In Fig.~4 the power law fit to the
equilibrium ($\dot\gamma=0$) relaxation times is reported as full
line, together with the relaxation times for three selected shear
rates: $\dot{\gamma}_1$=1$\cdot$10$^{-1}$ (full diamonds),
$\dot{\gamma}_2$=4$\cdot$10$^{-2}$ (full circles) and
$\dot{\gamma}_3$=6$\cdot$10$^{-3}$ (open circles). In the same
figure are also indicated by dashed lines the crossover
temperatures derived from Fig.~1 for the three selected shear
rates. Similarly to the thermodynamic quantities, also the
dynamics follow a simple behavior: at a fixed 
$\dot\gamma$ value, for
$T$ larger than a certain threshold the dynamics of the sheared
system is undistinguished from the equilibrium one. Below the
threshold the relaxation time flattens and no longer follows 
the steep increase 
associate to the slowing down of the dynamics which precedes the
glass transition. It is important
to emphasize that the crossover temperature, defined from the
{\it dynamics} is found to be the same as that derived from
{\it thermodynamic} data.

The existence of a well defined crossover temperature
$T_{\dot\gamma}$, as evidenced by the above results, suggests the
following scenario: for $T>T_{\dot\gamma}$ the sheared system is not
affected by the shear flow because the structural relaxation process
of the equilibrium system acts on a time-scale ($\tau_\alpha$) that is
faster than the one introduced by the shear flow; on lowering the
temperature $\tau_\alpha$ increases and -for temperatures close to
$T_{\dot\gamma}$- becomes comparable to the time scale introduced by
the shear (proportional to the inverse of shear rate
$\dot{\gamma}^{-1}$).  For $T<T_{\dot\gamma}$  
the shear starts to modify
both the static and the dynamic properties of the system. 
In the following, we  refer to the high temperature regime
($T>T_{\dot\gamma}$) as {\it ``$\alpha$-dominated''} 
({\it equilibrium} region)
and the low temperature regime ($T<T_{\dot\gamma}$) as {\it ``shear-dominated''}. 
Below $T_{\dot\gamma}$, in the {\it ``shear-dominated''} region,
the system, still remaining in a stationary state, ``freezes'' 
(in the sense that 
relevant processes governing the relaxations 
become those induced by
the shear and then only weakly temperature dependent). 
Dynamic behavior are
much less affected by further decreases of temperature and the
relaxation time reaches a finite value 
for $T\rightarrow$0.  
The previous scenario implies the existence of a strong 
relation between the equilibrium relaxation time at the 
crossover temperature and $\dot\gamma^{-1}$.  
In Fig.~5 we compare $\tau_{\alpha}(T)$ plotted as a function of
the temperature and
the inverse shear rate $\dot{\gamma}^{-1}$ 
plotted as a
function of $T_{\dot\gamma}$. The inverse
shear data have been multiplied by a constant factor $0.07$ 
in order to align them with the $\tau_{\alpha}(T)$ data. 
The temperature
dependence of the two quantities are in quite good agreement, 
suggesting a direct proportionality between them
\begin{equation}
\tau_\alpha (T_{\dot\gamma}) \propto \dot{\gamma}^{-1} \ .
\end{equation}
There is obviously a shear-independent prefactor that depends on the
chosen definition of $\tau_\alpha$ and on the specific investigated
correlation function (for example, 
the self relaxation time scale approximately as $q^{-2}$).
It is interesting to note the existence of a simple linear 
relationship between $\tau_\alpha$ and $\dot{\gamma}^{-1}$ along the line 
$T_{\dot \gamma}$ in the ($T$-$\dot\gamma$) plane,
as compared, for example, to the $\dot{\gamma}^{-2/3}$ dependence 
observed along the $T$-constant line \cite{bebaku00}.

Recent important work on the 
fluctuation-dissipation relation (FDR) 
in sheared  system \cite{beba,bebaku00,babe00}
has provided evidence that two different temperatures
control the dynamics.  Dynamics at short
times is controlled by the bath temperature, while
dynamics at longer times is controlled by an
effective temperature, larger than the bath one.
Our analysis predicts that only for $T<T_{\dot\gamma}$ 
a two-temperature scenario in the FDR should be observed.

Furthermore, our analysis confirms \cite{cukupe97} 
that an interesting and promising (from a numerical and
experimental point of view) implication of the above 
scenario could be the possibility to reproduce the 
aging properties of the non-driven system from the 
investigation of the shear effects on the
driven system. The underling main hypothesis 
is the similarity between the properties of the 
non-driven system at a given waiting time
and those of the driven system at a given shear rate
$\dot{\gamma}$.

In conclusion we have numerically studied the shear flow effects
on the thermal and dynamical properties of a simple model liquid,
focusing on the differences between driven and equilibrium
system, 
both in the dynamics and thermodynamics.
The potential energy and the
incoherent scattering function (and the associated structural
relaxation time) have been 
studied as a function of temperature for
different shear rates. 
It emerges the existence, for a given shear
rate, of a crossover line $T_{\dot\gamma}$, separating two
regimes: an high temperature (low shear rate)
regime in which the driven system behaves very similar to the
equilibrium one, and a low temperature (high shear rate) regime 
in which the driven system strongly deviate from the equilibrium
one. 
From the temperature dependence of relaxation times we found further 
evidences that $T_{\dot\gamma}$ marks the temperature at which the 
shear relaxation times start to deviate from the equilibrium times, 
passing from an {\it ``$\alpha$-dominated''} to a 
{\it ``shear-dominated''} region and approaching a finite value 
at low temperature.
Along the line defined by $T_{\dot \gamma}$ in the ($T$-$\dot\gamma$)
plane we observe a direct proportionality between $\tau_{\alpha}$ 
and $\dot{\gamma}^{-1}$, allowing a more clear interpretation of
the relationship between shear and relaxation times.

We acknowledge support from INFM Initiative Parallel Computing,
and MURST COFIN2000.

%%%%%%%%%%%%%%%%%%%%%%%%%%%%%%%%%%%%%%%%%%%%%%%%%%%%%%%%%%%%%%%%%%%%%%%%%%%
%                             REFERENCES
%%%%%%%%%%%%%%%%%%%%%%%%%%%%%%%%%%%%%%%%%%%%%%%%%%%%%%%%%%%%%%%%%%%%%%%%%%%

\clearpage
\newpage

\begin{figure}[t]
\centering
\vspace{.1cm}
\includegraphics[width=.8\textwidth,angle=0]{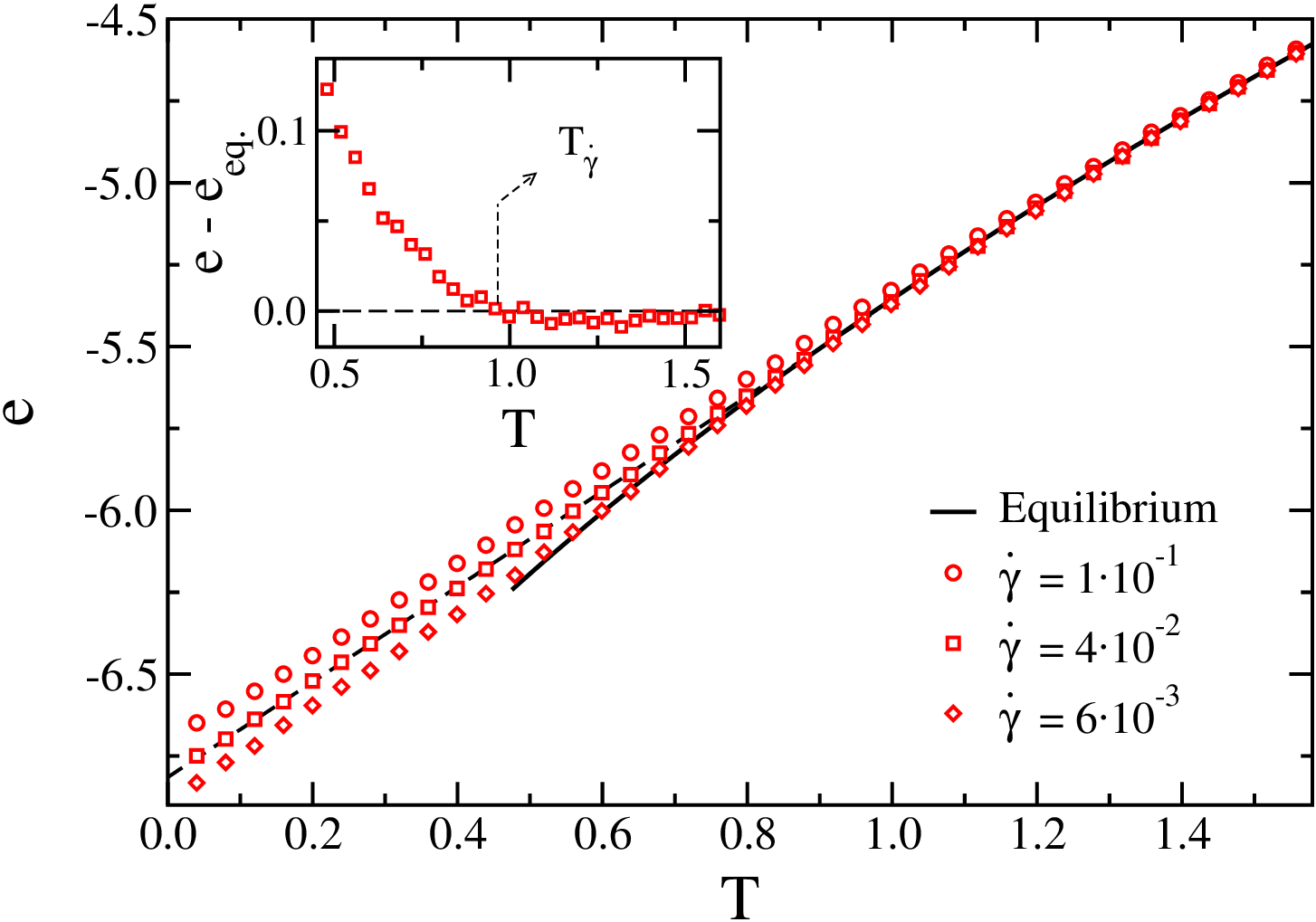}
\vspace{.1cm} \caption{Potential energy per particle $e$ as a function of
temperature for the equilibrium system (full line) and for driven
systems with different shear rates (symbols - from top to bottom
decreasing $\dot{\gamma} = 1 \cdot 10^{-1}, 4 \cdot 10^{-2},
6\cdot 10^{-3}$). The full line extends down to the lowest
temperature that we are able to equilibrate in the simulation. The
dashed line is the linear fit to the low temperature points of the
$\dot{\gamma} = 4 \cdot 10^{-2}$ curve. In the inset the
difference between the energy curve for the shear $\dot{\gamma} =
4 \cdot 10^{-2}$ and the $T^{3/5}$ fit of the equilibrium energy
$e_{eq.}$ as a function of $T$. The estimated crossover
temperature $T_{\dot\gamma}$ is also indicated by the arrow. }
\label{fig_1}
\end{figure}

\clearpage
\newpage

\begin{figure}[t]
\centering
\vspace{.05cm}
\includegraphics[width=.8\textwidth,angle=0]{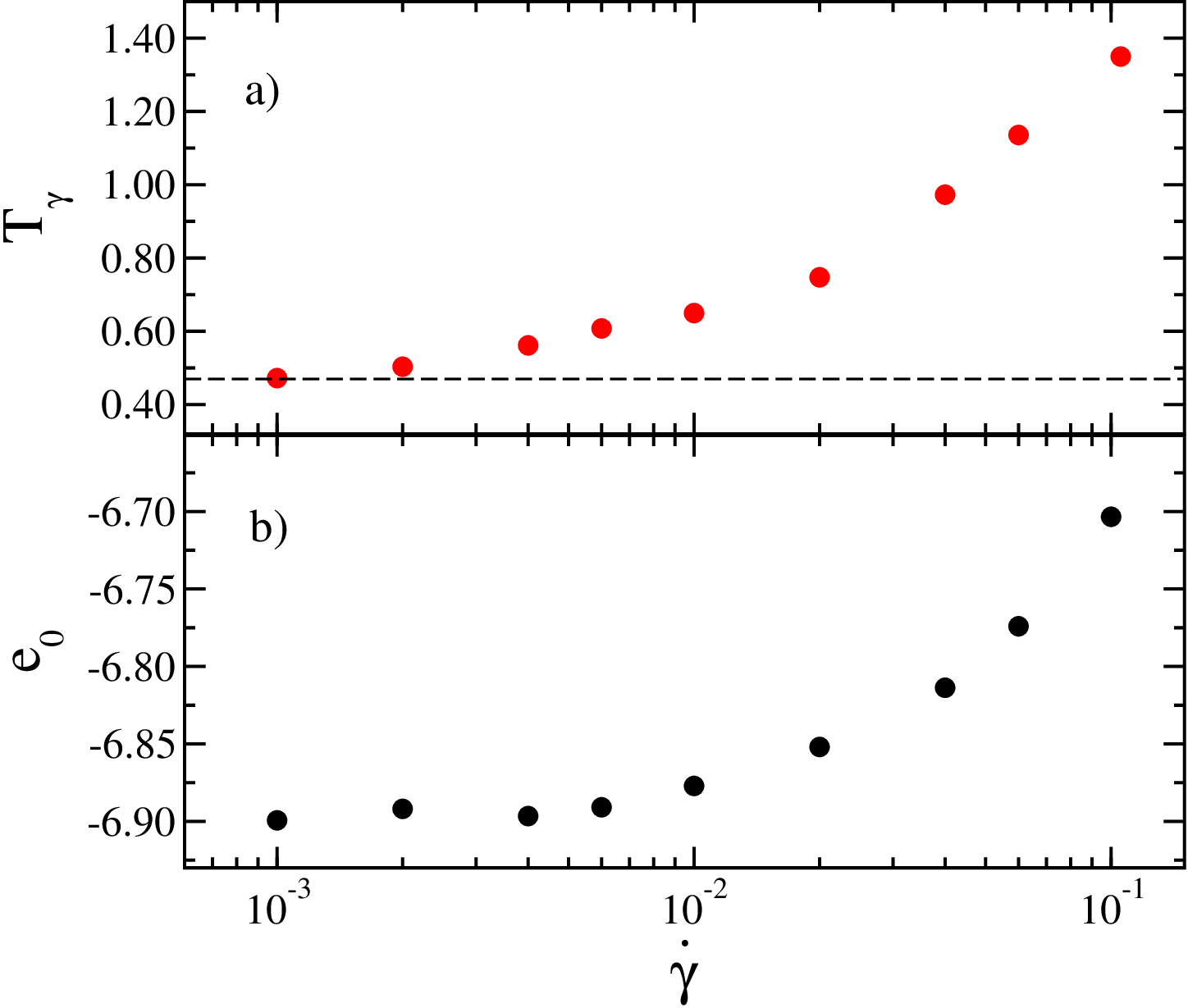}
\vspace{.1cm}
\caption{The shear rate dependence of the parameters describing
the energy curves of Fig. 1:
a) crossover temperature $T_{\dot\gamma}$ (intersection between low energy 
linear behavior and $T^{3/5}$ high temperature dependence in Fig. 1 
- the dashed line is the estimated value of the mode-coupling 
temperature $T_{MCT} \sim 0.475$ for this system),
b) zero temperature energy value $e_0$ (intersection of energy curves with the
$y$ axis in Fig. 1).
}
\label{fig_2}
\end{figure}

\clearpage
\newpage

\begin{figure}[t]
\centering
\vspace{.05cm}
\includegraphics[width=.8\textwidth,angle=0]{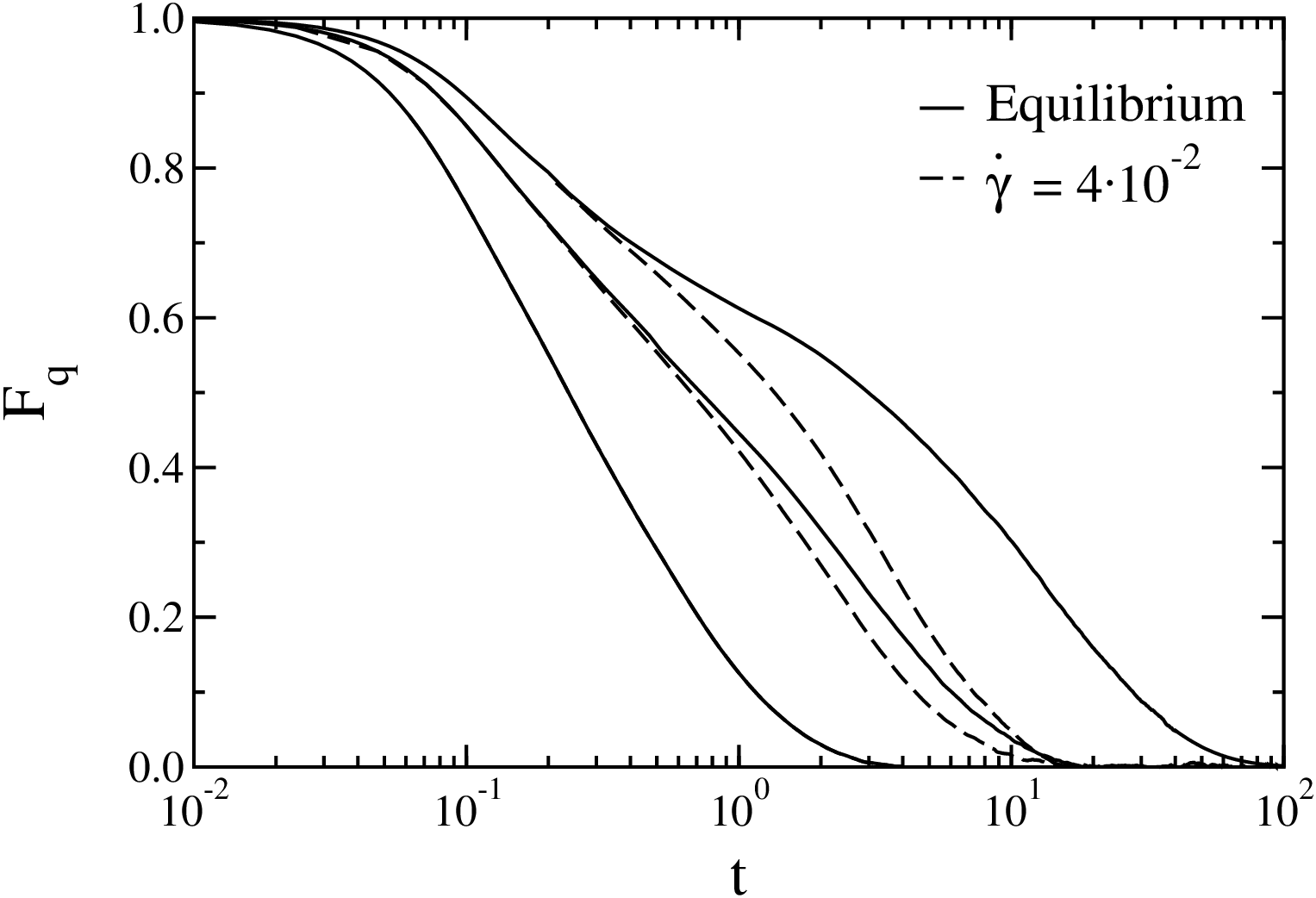}
\vspace{.1cm} \caption{Incoherent scattering functions $F_{\bf q}$
(calculated along the ``shear-free'' direction $z$ and for the
${\bf q}$ vector corresponding to the first peak of the static
structure factor, $q_{max}=7.1$) for the equilibrium system (full lines) and for
the driven systems with shear rate $\dot{\gamma} =  4 \cdot
10^{-2}$. The three curves refer to different temperatures: from
left to right $T = 1.6, 0.80, 0.56$ (we note that the crossover
temperature $T_{\dot\gamma}$ as defined from Fig. 1 for the shear rate
$\dot{\gamma} =  4 \cdot 10^{-2}$ is $T_{\dot\gamma} = 0.97$).
}\label{fig_3}
\end{figure}

\clearpage
\newpage

\begin{figure}[t]
\centering
\vspace{.05cm}
\includegraphics[width=.8\textwidth,angle=0]{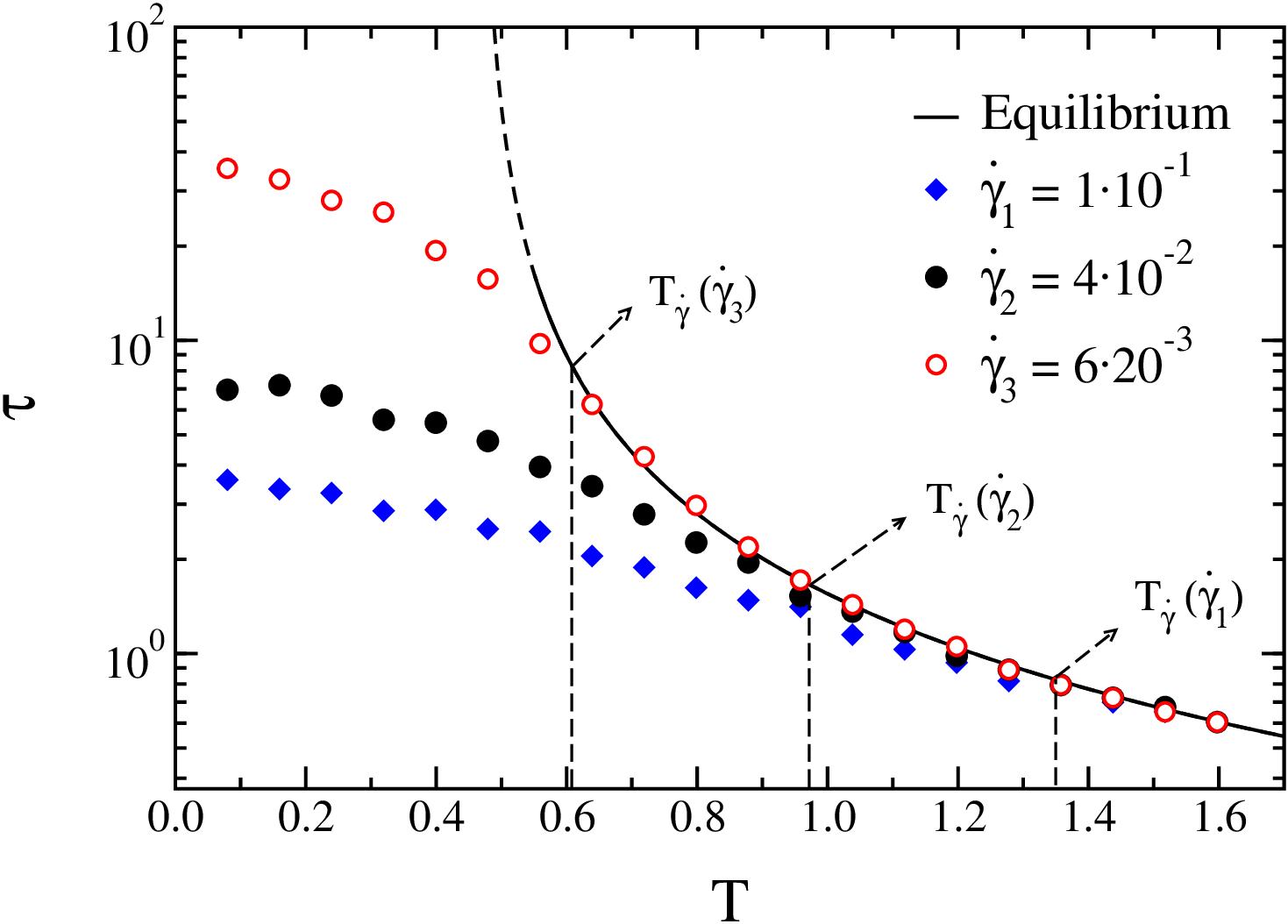}
\vspace{.1cm} \caption{Relaxation times (from incoherent scattering functions
at $q_{max}=7.1$, see Fig. 3) as a function of
temperature for the equilibrium system (the full line is a power
law fit of simulation data and dashed line is an extrapolation
below the last simulation point) and for the driven systems with
shear rates $\dot{\gamma}_1 =  1 \cdot 10^{-1}$ (full diamonds),
$\dot{\gamma}_2 =  4 \cdot 10^{-2}$ (full circles) and
$\dot{\gamma}_3 =  6 \cdot 10^{-3}$ (open circles). The arrows
indicate the crossover temperatures $T_{\dot\gamma}$ as defined in Fig.
1 for the corresponding shear rates: $T_{\dot\gamma}
(\dot{\gamma}_1)=1.35$, $T_{\dot\gamma} (\dot{\gamma}_2)=0.97$ and
$T_{\dot\gamma} (\dot{\gamma}_3)=0.61$. } \label{fig_4}
\end{figure}

\clearpage
\newpage

\begin{figure}[t]
\centering \vspace{.05cm}
\includegraphics[width=.8\textwidth,angle=0]{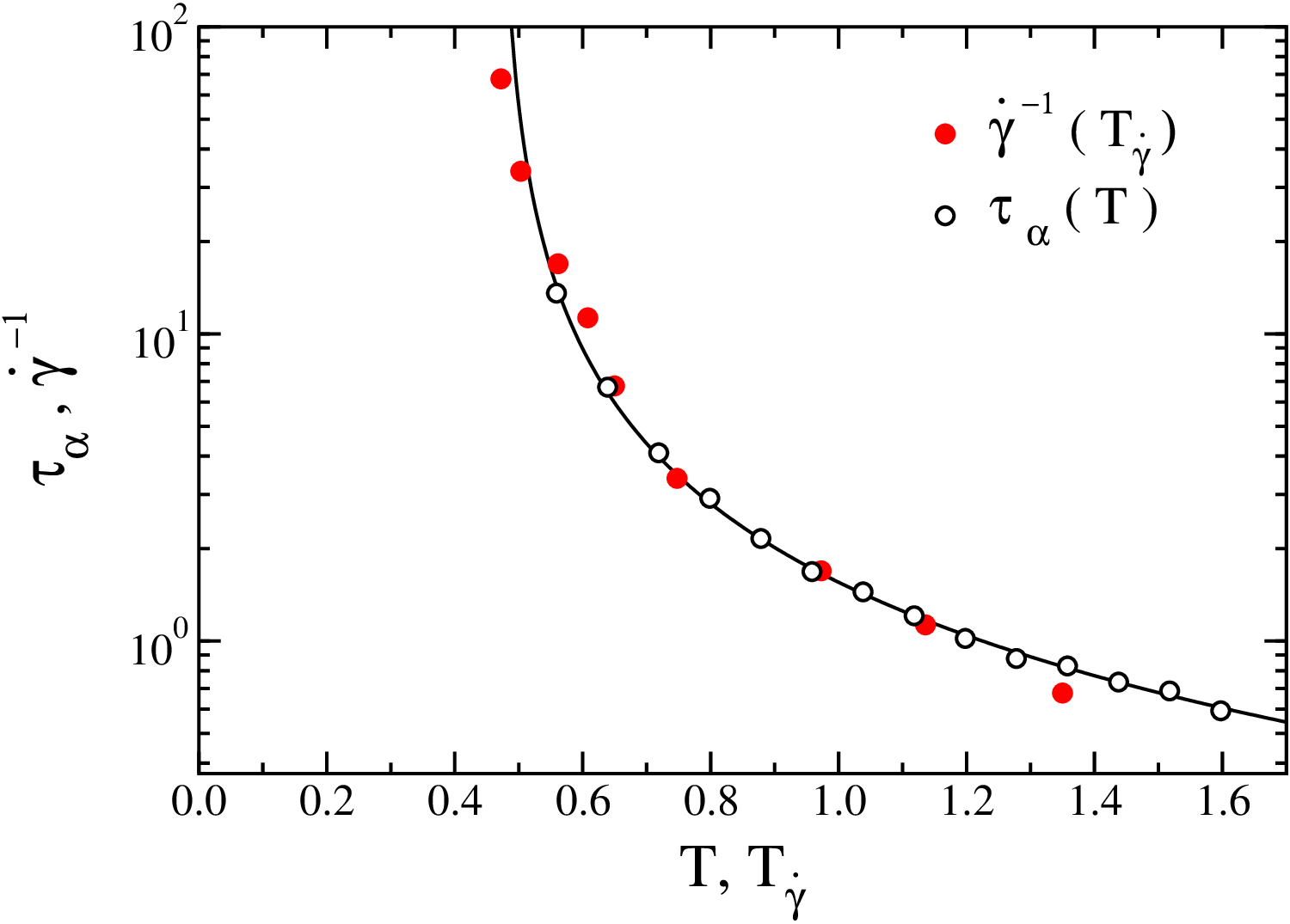}
\vspace{.1cm} \caption{Relaxation times $\tau_{\alpha}$ of the
equilibrium system (open symbols) as a function of temperature
(the full line is a power law fit), and the inverse shear rate
$\dot{\gamma}^{-1}$ (full symbols - multiplied by an arbitrary
factor $0.07$) as a function of crossover temperature $T_{\dot\gamma}$
as defined from Fig. 1. }\label{fig_5}
\end{figure}

\end{document}